\documentclass{INTERSPEECH2023}

\usepackage{array,booktabs}
\usepackage{color}
\usepackage{cite}
\usepackage{makecell}
\usepackage{multirow}
\usepackage{multicol}
\usepackage{amsmath,graphicx}
\usepackage{amssymb}
\usepackage{booktabs}
\newcommand{\ie}{\textit{i}.\textit{e}.}

\usepackage{arydshln}
\usepackage{pifont}
\newcommand{\pmr}[1]{\scriptsize$\pm$#1}

\interspeechcameraready 

\title{Intelligible Lip-to-Speech Synthesis with Speech Units}
\name{Jeongsoo Choi, Minsu Kim, Yong Man Ro}
\address{School of Electrical Engineering, KAIST, South Korea}
\email{\{jeongsoo.choi, ms.k, ymro\}@kaist.ac.kr}

\begin{document}

\maketitle
 
\begin{abstract}
In this paper, we propose a novel Lip-to-Speech synthesis (L2S) framework, for synthesizing intelligible speech from a silent lip movement video. Specifically, to complement the insufficient supervisory signal of the previous L2S model, we propose to use quantized self-supervised speech representations, named speech units, as an additional prediction target for the L2S model. Therefore, the proposed L2S model is trained to generate multiple targets, mel-spectrogram and speech units. As the speech units are discrete while mel-spectrogram is continuous, the proposed multi-target L2S model can be trained with strong content supervision, without using text-labeled data. Moreover, to accurately convert the synthesized mel-spectrogram into a waveform, we introduce a multi-input vocoder that can generate a clear waveform even from blurry and noisy mel-spectrogram by referring to the speech units. Extensive experimental results confirm the effectiveness of the proposed method in L2S.

\end{abstract}

\section{Introduction}
When the human voice is produced, both audio and visual signals are simultaneously produced as they are both from the same articulators. Therefore, speech audio and lip movement are highly correlated, meaning that linguistic information can be inferred not only from the audio but also from the visual modality. This correlation has led to the development of lip-reading \cite{Chung_2017_CVPR, martinez2020lipreading, kim2021cromm, prajwal2022sub}, also known as visual speech recognition, which aims to understand speech by solely depending on lip movements. Lip-reading can be helpful when the audio signal is corrupted by noise or when the speaker is too far away to hear clearly. While the performance of lip-reading models has improved significantly with the great advance of deep learning, training these models requires large-scale text-labeled data. However, obtaining accurate text transcriptions can be burdensome, leading to a scarcity of large video-text paired datasets in the public.

One possible solution for this problem is using Lip-to-Speech synthesis (L2S) technology, which directly generates spoken speech from silent lip video sources. This task can resolve the limitations of lip-reading by training an L2S model on audio-visual data which are more easily available than video-text paired data. However, developing L2S is regarded as a challenging task, since the mapping of video-to-audio is not one-to-one but one-to-many. This is because of the existence of homophenes, having the same mouth shape but with different pronunciations. In addition, since individuals have different voice characteristics, diverse speech can be produced from the same lip movements, according to speakers. Therefore, for developing an accurate L2S model, 1) fine-grained context modeling is necessary to distinguish the homophenes and 2) injecting speaker characteristics is required during synthesizing the speech from lip movements. Due to the challenges, previous works \cite{Prajwal_2020_CVPR, kim2021lip, hong2021speech} have mainly developed L2S methods by focusing on word-level or in-the-lab filmed datasets. Recently, several works\cite{schoburgcarrillodemira22_interspeech, kim2023lip} have attempted to develop L2S models on datasets constructed in the wild. \cite{schoburgcarrillodemira22_interspeech} proposed to use Conformer \cite{gulati20_interspeech} for modeling the context and to inject speaker embedding for modeling the speaker-specific information. However, even if they synthesized the speech from the wild datasets, the generated speech does not fully contain the original speech content of the lip movements. This is because they only utilize reconstruction loss for acoustic features (\ie, mel-spectrogram) which might be insufficient to guide the model focus on modeling the content. To address the limitation, \cite{kim2023lip} proposed to utilize additional supervision by using text. Through multi-task learning of speech reconstruction and speech recognition, they successfully generated speech with accurate content and improved the Word Error Rate (WER) of the synthesized speech. However, as they require text labels to provide content supervision, the video-text paired data should be employed to develop the L2S model, which might lose the benefit of the original L2S that is trainable without labeled data.

\begin{figure*}[t]
  \centering
  \includegraphics[width=\linewidth]{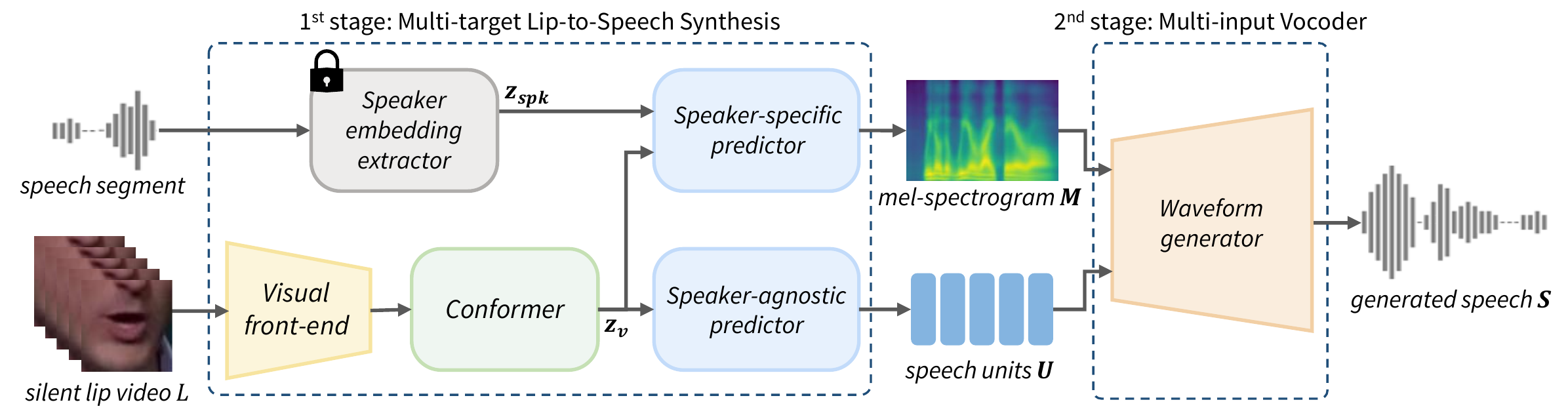}
  \caption{Overview of the proposed L2S framework. In the first stage, the proposed L2S model predicts mel-spectrogram along with speech units. By predicting speech units with mel-spectrogram, we can model more accurate speech content when predicting the speech from silent lip video. In the second stage, multi-input vocoder transforms the mel-spectrogram and the speech units into waveforms. We can generate intelligible waveforms from blurry predicted mel-spectrogram by utilizing additional input, the speech units.}
  \vspace{-0.3cm}
  \label{fig:1}
\end{figure*}

In this paper, we try to develop an intelligible L2S method that can synthesize speech with accurate content without using additional labels (\ie, text). Specifically, we employ quantized speech representations as pseudo-text, which can be obtained from input audio by using a speech model \cite{hsu2021hubert} pre-trained using Self-Supervised Learning (SSL). Previous works, GSLM \cite{lakhotia2021generative} and speech resynthesis \cite{polyak21_interspeech}, empirically have shown that quantizing the self-supervised speech representation can leave out the paralinguistic information and mainly keeps the linguistic information. Motivated by this, we utilize the quantized speech representations as pseudo-text, and we propose a novel multi-target L2S model that predicts mel-spectrogram and the quantized speech representations. We denote the quantized speech representations as `speech units' in the rest of the paper for brevity. By augmenting the L2S task with the prediction of speech units, the proposed L2S model can focus more on modeling content in the output speech without any additional labels.

In addition, we propose a novel multi-input vocoder, which is for converting the synthesized mel-spectrogram into waveform \cite{griffin1984signal, yamamoto2020parallel, kong2020hifi}. Since synthesized mel-spectrogram using L2S is usually blurry and noisy compared to the ground truth, directly utilizing a pre-trained vocoder that is trained on clean mel-spectrogram might produce unintelligible and noisy waveforms. In order to handle this, the proposed multi-input vocoder takes additional speech units as input which can make the vocoder focus on modeling accurate content in the waveform while other characteristics of speech (\ie, tones) can be inferred from the mel-spectrogram. Moreover, to reduce the gap between the train and test data, we augment the input mel-spectrogram with blur and noise during training. Therefore, the vocoder can learn to generate a high-fidelity waveform from the synthesized mel-spectrogram by referring to the speech units.

Our contributions are three-fold: 
(1) We propose a novel L2S method that can accurately model the speech content in the output speech by using quantized SSL speech representation as an auxiliary prediction target. 
(2) We design a multi-input neural vocoder that can generate intelligible waveforms from blurry and noisy mel-spectrogram by referring to the speech units. 
(3) The proposed method achieves better intelligibility scores of the generated speech compared to the previous methods.

\section{Proposed Approach}
\subsection{Overview}
When a silent talking face video is given, the proposed L2S framework aims to synthesize the speech that is synchronized with the input video. The proposed framework consists of two stages, as shown in Figure \ref{fig:1}. In the first stage, a multi-target L2S model predicts two speech features, mel-spectrogram and speech units, from a silent video. The speaker-specific characteristics such as pitch and amplitude are modeled along with the speech content in the output mel-spectrogram. To guide the L2S model to focus on modeling accurate speech content, we additionally predict speech units that mainly hold the linguistic information and speaker-agnostic. In the second stage, the predicted speech features are transformed into a waveform by using the proposed multi-input vocoder.

\subsection{Multi-target Lip-to-Speech Synthesis}
Synthesizing intelligible speech with accurate content from lip movements is challenging due to the absence of strong guidance for the content. Even if recent work \cite{kim2023lip} showed that providing text supervision to the model can boost the intelligibility of generated speech, it has one clear drawback of the necessity of labeled data, which might weaken the advantage of L2S tasks. In order to synthesize intelligible speech without the labeled data, we propose to utilize quantized SSL speech representation to represent linguistic information. Prior work \cite{lakhotia2021generative,inaguma2022unity,nguyen2023generative} showed that these discrete units have less paralinguistic information and mainly retain linguistic content information. Therefore, by using the speech units, we can provide frame-level content supervision to the L2S model even without using additional labels. The speech units can be obtained by quantizing speech representations (\ie, K-means) extracted from a pre-trained SSL representation model.

Specifically, when an input silent lip video $L$ is given, a visual front-end composed of ResNet-18 embeds the video input into visual features. The visual features are interleaved temporally following a constant ratio to match the length with the target speech. Then, Conformer encodes the contextualized features $z_v$ from the interleaved visual features. From the contextualized features, the model predicts two speech features, mel-spectrogram $M$ using a speaker-specific predictor and speech units $U$ using a speaker-agnostic predictor. Especially, in order to inject the speaker-specific information into the generated mel-spectrogram, we provide a speaker embedding $z_{spk}$ to the speaker-specific predictor. Following \cite{schoburgcarrillodemira22_interspeech}, we use \textit{d}-vector \cite{variani2014deep} for the speaker embedding which is obtained by feeding a speech segment into a pre-trained speaker verification model \cite{wan2018generalized}. Generating speech features from lip video can be represented as follows,
\begin{align}
    \hat{M}, \hat{U}=\mathcal{F}(L, z_{spk}),
\end{align}
where $\mathcal{F}(\cdot)$ represent the proposed multi-target L2S model, and $\hat{M}$ and $\hat{U}$ are the predicted mel-spectrogram and speech units, respectively.

In order to guide the proposed L2S model to generate speech feature, we employ L1 reconstruction loss to the predicted mel-spectrogram as follows,
\begin{align}
\mathcal{L}_{mel} = {||}M-\hat{M}{||}_1.
\end{align}
Moreover, to guide the model to focus on modeling accurate content during predicting mel-spectrogram, we apply frame-level content supervision using cross-entropy as follows,
\begin{align}
\mathcal{L}_{unit} = \sum_{t=1}^{T_a}U_t\log p(\hat{U}_t),
\end{align}
where $T_a$ is the length of speech units. By bringing the discrete representation prediction in the L2S task, which has been usually performed through the reconstruction of continuous acoustic features, we can provide more powerful supervision to the L2S model. The total loss is the weighted summation of the two loss functions, $\mathcal{L}_{tot} = \lambda_{mel}\mathcal{L}_{mel} + \lambda_{unit}\mathcal{L}_{unit}$, where $\lambda_{mel}$ and $\lambda_{unit}$ are the balancing weights for the loss functions.

\begin{table}[t]
  \renewcommand{\arraystretch}{1.2}
  \renewcommand{\tabcolsep}{3.0mm}
  \centering
  \caption{Experiments on LRS3 Dataset}
  \vspace{-0.2cm}
  \label{tab:table1}
  \resizebox{0.999\linewidth}{!}{
  \begin{tabular}{ c c c c c c }
    \Xhline{3\arrayrulewidth}
    \textbf{Method} & \textbf{STOI} & \textbf{ESTOI} & \textbf{PESQ} & \textbf{WER}\((\%)\) \\
    \hline
    VCA-GAN \cite{kim2021lip} & 0.474 & 0.207 & 1.23 & 95.9 \\
    SVTS \cite{schoburgcarrillodemira22_interspeech} & 0.507 & 0.271 & 1.25 & 78.0 \\
    Multi-Task \cite{kim2023lip}  & 0.496 & 0.266 & \textbf{1.31} & 65.8 \\
    \textbf{Proposed} & \textbf{0.543} & \textbf{0.351} & 1.28 & \textbf{50.2} \\ \hdashline
    \makecell{\textbf{Proposed} \\ \textbf{+ AV-HuBERT}} & 0.578 & 0.393 & 1.31 & 29.8 \\
    \Xhline{3\arrayrulewidth}
  \end{tabular}}
\vspace{-0.3cm}
\end{table}

\subsection{Multi-input Vocoder} 
\label{vocoder}
After predicting the speech features from the silent talking face video, we should transform the speech features into a waveform in order to hear the sound. Since the predicted mel-spectrogram is usually blurry and noisy compared to ground-truth, the content information can be lost during waveform conversion. To improve the intelligibility of the generated waveform from mel-spectrogram, we propose to additionally provide speech units to the vocoder. To this end, we design a multi-input vocoder $\mathcal{G}$ that takes both mel-spectrogram and speech units as inputs. We modified a GAN-based neural vocoder \cite{kong2020hifi} to allow the vocoder to generate waveforms that are conditionally dependent on additional linguistic content information, the speech units. Specifically, a lookup table is introduced to get an embedding according to each speech unit. Then, the embeddings are concatenated with the generated mel-spectrogram to be transformed into waveforms by the vocoder. Furthermore, in order to simulate the blurry and noisy mel-spectrogram synthesized by the L2S model, we propose to augment the input mel-spectrogram with blur and noise during training the vocoder. With this, we can enhance the model to accurately capture content information by referring to speech units and to produce an intelligible waveform, even from an erratic mel-spectrogram.

\begin{table}[t]
  \renewcommand{\arraystretch}{1.2}
  \renewcommand{\tabcolsep}{3.0mm}
  \centering  
  \caption{Experiments on LRS2 Dataset}
  \vspace{-0.2cm}
  \label{tab:table2}
  \resizebox{0.999\linewidth}{!}{
  \begin{tabular}{ c c c c c c }
    \Xhline{3\arrayrulewidth}
    \textbf{Method} & \textbf{STOI} & \textbf{ESTOI} & \textbf{PESQ} & \textbf{WER}\((\%)\) \\
    \hline
    VCA-GAN \cite{kim2021lip} & 0.407 & 0.134 & 1.24 & 101.1 \\
    SVTS \cite{schoburgcarrillodemira22_interspeech} & 0.530 & 0.331 & 1.30 & 71.4 \\
    Multi-Task \cite{kim2023lip}  & 0.526 & 0.341 & \textbf{1.36} & 57.8 \\ 
    \textbf{Proposed} & \textbf{0.565} & \textbf{0.395} & 1.32 & \textbf{44.8} \\ \hdashline
    \makecell{\textbf{Proposed} \\ \textbf{+ AV-HuBERT}} & 0.585 & 0.412 & 1.34 & 35.7 \\
    \Xhline{3\arrayrulewidth}
  \end{tabular}}
\vspace{-0.3cm}
\end{table}

\vspace{-0.1cm}
\section{Experiments}
\subsection{Datasets}

\noindent \textbf{LRS3} \cite{afouras2018lrs3} dataset is a large-scale sentence-level dataset obtained from TED and TEDx videos, which makes 433 hours of the audio-visual corpus of thousands of speakers. Since the dataset is obtained from TED, the video contains natural pose and illumination variations.
We follow the data splits of \cite{schoburgcarrillodemira22_interspeech} to simulate the unseen speaker test setting.

\noindent \textbf{LRS2} \cite{afouras2018deep} dataset is another large-scale dataset comprising 223 hours of transcribed audio-visual corpus which are filmed in the wild environment, sourced from BBC. There are thousands of speakers but no speaker information is provided.

\subsection{Implementation Details} \label{implementation}
\subsubsection{Preprocessing}
We utilize dlib\cite{king2009dlib} to detect 68 facial landmarks in every frame of each video and align faces to a reference face. We then crop the lip region as a size of 96$\times$96 patch and convert it into grayscale. The lip video is center-cropped into a size of 88$\times$88 during inference. 
We resample the audio to 16kHz and extract mel-spectrogram using 80 mel filter banks, 40ms window, and 10ms hop size, resulting in 100Hz. In order to extract speech units, we leverage a publicly available SSL speech representation model, HuBERT-BASE\cite{hsu2021hubert}, and K-means clustering model which is trained on LJ Speech\cite{ljspeech17} dataset. The number of clusters (\ie, codebook size) of the speech units is set to 200 and the units are sampled at 50Hz. 20ms of mel-spectrogram is stacked to match the sampling rate of speech units when used as the target of the L2S model and unstacked before passing through the multi-input vocoder.

\subsubsection{Architecture}
For fair comparisons, we follow the previous works \cite{schoburgcarrillodemira22_interspeech, kim2023lip} which have Conformer encoder with 12 layers, 8 attention heads, convolutional kernel size of 31, and a latent dimension of 512. The input of speaker-specific predictor is the concatenated feature of contextualized visual feature $z_v$ and the speaker embedding $z_{spk}$. We utilize 1D CNN for the speaker-specific predictor and MLP for the speaker-agnostic predictor. The architecture of the multi-input vocoder is modified from HiFi-GAN \cite{kong2020hifi} and the inputs are mel-spectrogram, speech units, and speaker embedding. The speech units are embedded into 128-dimensional vectors and temporally upsampled to have the same length as mel-spectrogram. The speaker embedding is also projected into a 128-dimensional vector and upsampled by repetition.

\subsubsection{Training Details} \label{detail}
When training the multi-target L2S model, we randomly crop the lip video to a spatial size of 88$\times$88 and horizontally flip it with a probability of 0.5. During inference, the lip video is center-cropped and no horizontal flipping is performed. Random erasing and time masking to the cropped lip video are utilized for data augmentation as described in \cite{schoburgcarrillodemira22_interspeech} during training. We set $\lambda_{mel}$ and $\lambda_{unit}$ to 10 and 1, respectively. We train the model for 150 epochs on each dataset using Fairseq\cite{ott2019fairseq} with Adam\cite{kingma2014adam} optimizer and learning rate of $10^{-3}$ with cosine schedule. When AV-HuBERT LARGE encoder is utilized for the visual front-end, we only train the model 50 epochs by freezing AV-HuBERT. We separately train the multi-input vocoder on each dataset for 1M updates with a batch size of 16 using a single A6000 GPU. When training with augmentation, we regard the log mel-spectrogram as a grayscale image, and apply Gaussian blur whose kernel size is 9$\times$9 with the standard deviation range of 0.1 to 1. We also add Gaussian noise with a maximum standard deviation of 1. 

\subsection{Evaluation Metrics}
We evaluate generated samples by each method using several metrics using the test set of each dataset. Short-Time Objective Intelligibility (STOI)\cite{taal2011algorithm} metric and Extended STOI (ESTOI) \cite{jensen2016estoi} metric are employed to measure the intelligibility of each generated sample. Perceptual Evaluation of Speech Quality (PESQ) is used to assess perceptual speech quality. In order to evaluate the intelligibility of the content, we use Word Error Rate (WER). WER is measured by performing speech recognition of the generated speech using a pre-trained model \cite{ma2021end} which is trained on the audio of each dataset. The WER using ground-truth audio is 4.2\% for LRS2 and 2.5\% for LRS3, respectively. Higher STOI, ESTOI, and PESQ, and lower WER indicate better-generated speech.

\begin{table}[t]
  \renewcommand{\arraystretch}{1.2}
  \renewcommand{\tabcolsep}{2.5mm}
  \centering
  \caption{Ablation Study on LRS3 Dataset}
  \vspace{-0.2cm}
  \label{tab:table3}
  \resizebox{0.999\linewidth}{!}{
  \begin{tabular}{ l c c c c }
    \Xhline{3\arrayrulewidth}
    \textbf{Method} & \textbf{STOI} & \textbf{ESTOI} & \textbf{PESQ} & \textbf{WER(\%)} \\ \hline
    Baseline \cite{schoburgcarrillodemira22_interspeech} & 0.516 & 0.292 & 1.27 & 72.5 \\
    $\quad$ + Speech units & 0.542 & 0.343 & 1.29 & 50.7 \\
    $\quad$ + Multi-input vocoder & \textbf{0.552} & \textbf{0.354} & \textbf{1.31} & 50.4 \\
    $\quad$ + Augmented mel & 0.543 & 0.351 & 1.28 & \textbf{50.2} \\ \hdashline
    $\quad$ + AV-HuBERT \cite{shi2022learning} & 0.578 & 0.393 & 1.31 & 29.8 \\
    \Xhline{3\arrayrulewidth}
  \end{tabular}}
\vspace{-0.1cm}
\end{table}

\section{Results}
\subsection{Quantitative Comparison}
Table \ref{tab:table1} and \ref{tab:table2} show the evaluation results of the proposed methods and the previous methods on LRS3 and LRS2, respectively. The proposed L2S framework demonstrates superior performances in terms of STOI and ESTOI, indicating that it can reconstruct speech-related features with more detail than previous methods. Additionally, the lowest WER represents that the proposed L2S framework can synthesize authentic waveforms that convey plausible linguistic content. Compared to the previous state-of-the-art method, Multi-Task, which utilizes text transcription during training, the proposed method outperforms in terms of intelligibility, even though we do not use any text labels in training. We attribute this performance gain to the frame-level linguistic content supervision using speech units and the proposed multi-input vocoder. 
In addition, since the proposed method can be employed with the recent powerful representation model, AV-HuBERT \cite{shi2022learning,shi2022robust}, we additionally report the performance of the proposed method built on AV-HuBERT. We use the publicly available AV-HuBERT LARGE model which is pre-trained on LRS3\cite{afouras2018lrs3} and VoxCeleb2\cite{chung18b_interspeech} datasets. After replacing the visual front-end with the AV-HuBERT encoder for extracting visual features, the performance improves significantly on both datasets. In particular, the obtained WER performance is comparable to state-of-the-art lip-reading models \cite{kim2022distinguishing,ma2021end}, which demonstrates the superiority of the proposed method in generating intelligible speech from silent talking face videos. Please note that the proposed L2S framework does not require any text data, but still achieves comparable performance to the lip-reading model.

\begin{table}[t]
  \renewcommand{\arraystretch}{1.2}
  \renewcommand{\tabcolsep}{3.0mm}
  \centering
  \caption{MOS Comparison on LRS3 Dataset}
  \vspace{-0.2cm}
  \label{tab:table4}
  \resizebox{0.999\linewidth}{!}{
  \begin{tabular}{ c c c c c }
    \Xhline{3\arrayrulewidth}
    \textbf{Method} & \textbf{Naturalness} & \textbf{Intelligibility} & \textbf{Clearness} \\
    \hline
    VCA-GAN \cite{kim2021lip} & 1.24\pmr{0.12} & 1.64\pmr{0.33} & 1.25\pmr{0.18} \\
    SVTS \cite{schoburgcarrillodemira22_interspeech} & 1.98\pmr{0.25} & 2.35\pmr{0.27} & 1.95\pmr{0.24} \\
    Multi-Task \cite{kim2023lip} & 1.65\pmr{0.17} & 2.29\pmr{0.33} & 1.64\pmr{0.16} \\
    \textbf{Proposed w/o aug} & 3.28\pmr{0.27} & 3.38\pmr{0.16} & 3.03\pmr{0.24} \\
    \textbf{Proposed} & \textbf{4.40}\pmr{0.13} & \textbf{3.92}\pmr{0.15} & \textbf{4.15}\pmr{0.15} \\
    \hdashline
    Ground-truth      & 4.98\pmr{0.03} & 4.94\pmr{0.04} & 4.98\pmr{0.02} \\
    \Xhline{3\arrayrulewidth}
  \end{tabular}}
\vspace{-0.1cm}
\end{table}

\subsection{Ablation Study}
We conduct an ablation study to examine the effect of each component of the proposed methods. The ablation result is shown in Table \ref{tab:table3}. The baseline model is SVTS \cite{schoburgcarrillodemira22_interspeech} which is slightly modified to match our framework except for speech units. By using speech units and providing additional content supervision to the model, the performance is significantly improved indicating that the additional target provides valuable linguistic content and enhances the extraction of meaningful content from lip movement video. Furthermore, replacing the vocoder with the proposed multi-input vocoder results in more intelligible outputs by achieving improved STOI and ESTOI. Although the predicted speech units from the multi-target L2S model are not perfect, they can serve as an additional discrete condition for the vocoder to reduce artifacts and help the model generate high-fidelity waveforms. Finally, we confirm that our augmentation technique, introduced in Section \ref{vocoder}, further improves the ability of the framework to create clean speech. While metrics except WER show slight performance drops, the human evaluation that reflects how it indeed hears to humans indicates that the augmentation technique helps to generate high-quality waveforms. Notably, the technique also significantly reduces noise in the generated waveform as expected. Please refer to Section \ref{user_study} and listen to the provided speech samples in the supplemental material for accurate evaluation. Finally, by using a pre-trained visual encoder, AV-HuBERT, the overall performance is significantly improved. Especially, we achieve 29.8\% WER on LRS3, which is a significant result compared to the previous state-of-the-art method \cite{kim2023lip} that achieves 65.8\% WER.

\subsection{Human Evaluation} \label{user_study}
Finally, we conduct a human subjective test to evaluate the quality of the generated speech. A total of 15 participants are asked to rate their opinions on naturalness, intelligibility, and clearness on a scale of 1 (least) to 5 (most). For this user study, we use 20 randomly selected speech samples from the LRS3 test set and calculate mean opinion scores (MOS). Naturalness refers to how similar the speech is to that of a human, intelligibility measures how easily the linguistic content is understood, and clearness assesses the level of noise in the speech. The Larger score means a clearer sound. For the previous methods, we use test samples provided by the authors. Table \ref{tab:table4} presents a comparison of MOS between previous methods and our proposed method, indicating that our method is superior to the previous works in terms of various criteria for speech quality. Please note that we use the same visual encoder as the previous works for human evaluation, not AV-HuBERT. Furthermore, in order to verify the effectiveness of the proposed augmentation technique of mel-spectrogram in training the multi-input vocoder, we also report the MOS result for the generated speech without the augmentation. The human evaluation results clearly confirm that using the augmented mel-spectrograms for training vocoder improves both the naturalness and clearness of the generated speech.

\section{Conclusions}
In this paper, we propose a novel L2S framework for synthesizing intelligible speech, even without using text data. Our approach utilizes speech units to provide frame-level content supervision to L2S model, enabling improved content modeling ability. Additionally, we design a multi-input neural vocoder that takes both mel-spectrogram and speech units as inputs, ensuring the generation of high-fidelity waveforms even from blurry and noisy mel-spectrograms during inference. Experimental results show that the proposed method outperforms previous methods in terms of intelligibility and low-level feature reconstruction performance.

\section{Acknowledgements}
This work was supported by the National Research Foundation of Korea (NRF) grant funded by the Korea government (MSIT) (No.NRF-2022R1A2C2005529).

\bibliographystyle{IEEEtran}
\bibliography{main}

\end{document}